\documentclass[conference]{IEEEtran}

\usepackage[T1]{fontenc}
\usepackage[utf8]{inputenc}
\usepackage{amsmath, amssymb, amsfonts}
\usepackage{graphicx}
\usepackage{textcomp}
\usepackage{xcolor}
\usepackage{cite}
\usepackage{url}
\usepackage{booktabs}
\usepackage{multirow}
\usepackage{microtype}

\title{A Comparative Study of Pre-trained Speech Encoders and Training Objectives for Large-Scale Indic
Spoken Language Identification}

\author{
\IEEEauthorblockN{
Agneedh Basu\IEEEauthorrefmark{1},
Pavan Kumar J\IEEEauthorrefmark{1},
Sujith P\IEEEauthorrefmark{1},
Visruth Sanka\IEEEauthorrefmark{1},
Nihar Desai\IEEEauthorrefmark{1},
Prasanta Kumar Ghosh \IEEEauthorrefmark{2}
}

\IEEEauthorblockA{\IEEEauthorrefmark{1}
AI \& Robotics Technology Park (ARTPARK), IISc, Bangalore, India \\
}

\IEEEauthorblockA{\IEEEauthorrefmark{2}
Department of Electrical Engineering, Indian Institute of Science, Bangalore, India \\
}
}

\begin{document}

\maketitle

% ---- Abstract ---------------------------------------------
\begin{abstract}
Spoken language identification (LID) for Indian languages is a
challenging problem due to the large number of languages, significant
phonetic overlap among related varieties, and the scarcity of
labeled data for many low-resource languages. In this work, we
present a systematic comparative study of two pre-trained speech
encoders---Whisper and FastConformer---combined with a linear classifier
for large-scale Indic LID spanning 42 languages across four
linguistic families. We evaluate both encoders in frozen (linear
probing) and fine-tuned settings, and compare three training
objectives: cross-entropy (CE), supervised contrastive loss with cross entropy (CE + supCon), and hierarchical
softmax (HSM). Models are trained on the Vaani dataset and evaluated
in a cross-corpus setting on Vaani-Test (held-out), FLEURS, and Kathbath,
providing insights into domain generalization. The frozen FastConformer encoder
achieves over 90\% macro accuracy on FLEURS and Kathbath without
any task-specific adaptation, substantially outperforming Whisper
on out-of-domain benchmarks, while fine-tuned Whisper yields
stronger in-domain performance. HSM consistently outperforms CE
and CE+SupCon for both encoders across all benchmarks, with the
largest gains on out-of-domain test sets. CE+SupCon degrades
FastConformer's cross-corpus generalization, suggesting that the
contrastive objective over-specializes representations to in-domain
conditions. Per-family analysis shows that Central Indo-Aryan
varieties are the hardest to discriminate, with
Hindi--Urdu and the Sadri--Chhattisgarhi--Surgujia cluster
being the dominant confusion pairs.
\end{abstract}

% ---- Keywords ---------------------------------------------

% ===========================================================
\section{Introduction}
\label{sec:intro}

India is home to over 1,600 mother tongues~\cite{langcensus}, with 22 constitutionally recognized languages~\cite{officiallangs} spanning three major families: Indo-Aryan, Dravidian, Sino-Tibetan(\cite{masica1991indoaryan, abbi2006india}). Automatic spoken language identification (LID) is a critical front-end component for multilingual speech pipelines, enabling ASR, machine translation, and voice-based services to route input to the correct
language-specific model. Indic LID poses distinct challenges: closely related language
varieties -- such as Hindi, Bhojpuri, Chhattisgarhi, and Sadri -- exhibit strong phonological and lexical overlap,
making fine-grained discrimination difficult; many target
languages are low-resource, with limited training data available, existing benchmarks differ substantially in recording
conditions, speaking style, and speaker demographics, making
cross-corpus generalization a practical concern.

Recent advances in self-supervised and weakly supervised speech
representation learning have yielded powerful pre-trained
encoders~\cite{radford2023whisper, gulati2020conformer} that can be adapted
to downstream tasks with minimal supervision. However, it is
unclear which encoder architecture is better suited for Indic LID,
and how much the choice of training objective---standard
cross-entropy (CE), metric learning via contrastive loss, or
hierarchically-structured softmax---affects downstream performance
and generalization.

In this paper, we address these questions through a controlled
empirical study. Specifically, we make the following contributions:
\begin{itemize}
  \item We benchmark Whisper and FastConformer encoders, both frozen
        and fine-tuned, as feature extractors for Indic LID across
        42 languages.
  \item We compare three training objectives -- cross entropy, contrastive loss, and hierarchical soft max -- on the same encoder backbones and report results across three test benchmarks (Section~\ref{sec:losses}).
  \item We conduct a cross-corpus evaluation to assess domain generalization
        (Section~\ref{sec:results}).
  \item We provide a per-family error analysis to understand the
        role of linguistic proximity in LID confusion
        (Section~\ref{sec:analysis}).
\end{itemize}

% ===========================================================
\section{Related Work}
\label{sec:related}

Early deep learning approaches to Indic LID used log-Mel
spectrograms as input to CNN-based encoders with a linear
classification head. Godbole et al.~\cite{Godbole}
demonstrated over 98\% accuracy on seven Indian languages from the
IIIT-H Indic speech database using this pipeline, and a related
study~\cite{visualize_lid_2019} extended this to CNN+ANN classifiers
on the same corpus, reporting accuracies above 99\% while showing
that CNN feature maps produce interpretable language-specific
patterns. Despite these high numbers, both works evaluate on
in-domain data only, leaving cross-corpus generalization
unexplored. A survey of Indic LID~\cite{indic_lid_survey_2022} reviews
existing systems across 23 studies and highlights the limited
language coverage and dataset diversity of prior work. Self-supervised representations have more recently been applied to Indic LID. Sai et al.~\cite{phonotactic_ssl_2021} explored phonotactic representations for seven Indian languages using wav2vec-based self-supervised features and senone classifier hidden representations, combined with linear, BiLSTM, or CNN classifiers. Self-supervised features consistently outperformed supervised counterparts, achieving approximately 94\% accuracy. Cross-corpus evaluation has emerged as a key challenge for Indic LID. Bhatt et al.~\cite{cross_corpora_eusipco_2021} conducted a preliminary cross-corpus study across three Indian language corpora (IIITH-ILSC, LDC South Asian, IITKGP-MLILSC) using MFCC features with a TDNN architecture, showing that normalization via CMS and RASTA-filtering substantially improves out-of-domain LID performance. Extending this, \cite{cross_corpora_csl_2023}
investigated cross-corpus Indic LID using ECAPA-TDNN with MFCC
features, proposing domain diversification and generalization
strategies to reduce the train--test domain gap. Domain-adaptive objectives have also been explored. Bhatt et al.~\cite{centroid_slt_2021} combined centroid classification loss with cross-entropy for LID in unseen target domains, achieving 94\% on eight Indian languages.

Most closely related to our work, Shrivastava et
al.~\cite{shrivastava2024indicwhisper} fine-tuned Whisper-large-v3 for
joint ASR and LID on five Indian languages by incorporating
language tokens into the ASR token sequence during training,
reducing diarization error rate from 24.04\% (vanilla Whisper)
to 18.03\% on the DiSPLACE 2024 challenge. Unlike that work, we
study encoder representations in isolation as feature extractors
for standalone LID, evaluated on a much broader set of 42 Indian
languages across Vaani, FLEURS, and Kathbath. To our knowledge, LID has not been attempted on this scale of diverse Indian languages.

% \subsection{Training Objectives for Classification}
% Cross-entropy (CE) remains the dominant training objective for LID,
% directly optimizing the final classification layer. Supervised
% contrastive learning (SupCon)~\cite{khosla2020supcon} improves
% representation quality by pulling together same-class embeddings
% in the embedding space; however, it does not directly supervise the
% classification layer, making it unsuitable as a standalone objective
% for a classifier. Khosla et al.~\cite{khosla2020supcon} suggest combining
% SupCon with CE so that the embedding space is shaped by contrastive
% objectives while the classification head is simultaneously supervised.
% In this work, we adopt this combined objective and compare it against
% CE-only and hierarchical softmax (HSM)~\cite{morin2005hsm}.

% ===========================================================
\section{System Description}
\label{sec:system}

\subsection{Model Architecture}
Our system follows a simple encoder--classifier architecture: a pre-trained speech encoder maps an input utterance to a sequence of frame-level representations, which are aggregated via a pooling layer to obtain a fixed-length embedding vector $\mathbf{z}$. This embedding is then passed to a linear classification head. The encoder is either held frozen or fine-tuned end-to-end.

(a)Whisper: We use the encoder of \texttt{openai / whisper-medium}~\cite{radford2023whisper} which is trained for ASR on 90+ languages and produces a 1024-dimensional frame-level representation. This encoder alone comprises of 350M parameters. (b)FastConformer: We use the encoder from \texttt{ARTPARK-IISc / Vaani-FastConformer-Multilingual} ~\cite{vaani2023} which is trained on 60+ Indic languages, with an embedding dimension of 1024. The encoder consists of 430M parameters.

Self-Attention-pooling\cite{selfattention} over frames yields a fixed-length utterance embedding. A single linear layer projects the utterance embedding to logits over $L = 42$ target languages. In the \emph{frozen} setting, only
the final two layers are trained (linear probing). In the \emph{fine-tuned}
setting, encoder weights are updated jointly with these two layers.

\subsection{Training Objectives}
\label{sec:losses}

\paragraph{Cross-Entropy (CE)}
Standard multi-class cross-entropy:
\begin{equation}
  \mathcal{L}_{\text{CE}} = -\sum_{l=1}^{L} y_l \log \hat{p}_l,
  \label{eq:ce}
\end{equation}
where $y_l \in \{0,1\}$ is the ground-truth indicator for language
$l$ and $\hat{p}_l$ is the predicted softmax probability.

\paragraph{Combined CE + SupCon Loss (CE + SupCon)}
Standalone supervised contrastive loss (SupCon)~\cite{khosla2020supcon}
optimizes the embedding space but provides no direct gradient to the
classification head. Conversely, CE trains the output layer but does
not explicitly encourage discriminative clustering in the embedding
space. To benefit from both, we combine the two objectives:
\begin{equation}
  \mathcal{L}_{\text{CE+SupCon}} = \lambda\,\mathcal{L}_{\text{SupCon}}  + (1-\lambda)\mathcal{L}_{\text{CE}},
  \label{eq:combined}
\end{equation}
where $\lambda$ is a weighting hyperparameter. The SupCon term is:
\begin{equation}
  \mathcal{L}_{\text{SupCon}} = -\frac{1}{|P(i)|} \sum_{p \in P(i)}
  \log \frac{e^{\mathbf{z}_i \cdot \mathbf{z}_p / \tau}}
            {\sum_{a \neq i} e^{\mathbf{z}_i \cdot \mathbf{z}_a / \tau}},
  \label{eq:supcon}
\end{equation}
where $\mathbf{z}_i$ is the $\ell_2$-normalized embedding of sample
$i$, $P(i)$ the set of same-language positives in the batch, and
$\tau$ a temperature hyperparameter. The combined loss thus
simultaneously shapes the embedding space via $\mathcal{L}_{\text{SupCon}}$
and supervises the classification layer via $\mathcal{L}_{\text{CE}}$.

\paragraph{Hierarchical Softmax (HSM)}
We structure the 42 languages into a four-level linguistic tree:
Root $\to$ Family $\to$ Sub-family $\to$ Language, with families
Indo-Aryan, Dravidian, Sino-Tibetan, and European. Each internal node is associated with a linear classifier that predicts among its children;
the language probability is the product of branch probabilities
along the root-to-leaf path~\cite{morin2005hsm}. Let $\mathbf{z}_i \in \mathbb{R}^d$ denote the embedding for sample $i$, and $y_i$ its label. Let the path to class $y_i$ in the hierarchy be:
$ \mathcal{P}(y_i) = \{(n_1, c_1), \dots, (n_{L_i}, c_{L_i})\}$ where $n_\ell$ is a node and $c_\ell$ is the index of the correct child at that node. The conditional probability at node $n_\ell$ is 
$P(c_\ell \mid n_\ell, \mathbf{z}_i) = \mathrm{softmax}(W_{n_\ell} \mathbf{z}_i)_{c_\ell} $. The hierarchical likelihood is
\[
P(y_i \mid \mathbf{z}_i) = \prod_{\ell=1}^{L_i} P(c_\ell \mid n_\ell, \mathbf{z}_i)
\]

The hierarchical softmax loss is:
\[
\mathcal{L}_{\mathrm{HSM}} = - \sum_{\ell=1}^{L_i} 
\log P(c_\ell^{(i)} \mid n_\ell^{(i)}, \mathbf{z}_i)
\]

% ===========================================================
\section{Experimental Setup}
\label{sec:setup}

\subsection{Datasets}
Vaani~\cite{vaani2023} is a large-scale naturalistic Indic speech corpus. 
From this corpus, we curate a balanced subset by selecting 10 hours of speech from each of the 42 languages (see Appendix~\ref{app:languages} for the full list), maximizing diversity in districts and speakers. The resulting data is split into train, validation, and test 
sets in an 8:1:1 ratio, ensuring that speakers do not overlap across splits.

\paragraph{Evaluation}
We adopt a cross-corpus evaluation protocol: models are trained
exclusively on Vaani and tested on three benchmarks to assess
in-domain and out-of-domain performance -- \textbf{Vaani-Test}: held-out portion of Vaani (in-domain); 13 languages from \textbf{FLEURS}~\cite{conneau2023fleurs}; and  11 from \textbf{Kathbath}~\cite{javed2022kathbath}.

\subsection{Implementation Details}
All models are optimized using AdamW~\cite{loshchilov2019adamw} with linear learning rate warmup following~\cite{devlin2019bert}.
For full fine-tuning, we use a learning rate of $1\mathrm{e}{-5}$, while for linear probing, a higher learning rate of $1\mathrm{e}{-4}$ is used. 
Training is performed with a batch size of 4 and gradient accumulation over 24 steps, resulting in an effective batch size of 96. 
The combined CE + SupCon loss uses a temperature $\tau = 0.07$ and weighting $\lambda = 0.5$. All experiments are conducted on an NVIDIA L40 GPU.
\subsection{Evaluation Metrics}
We report macro-averaged accuracy (recall) over all languages
present in each test set. Macro averaging treats each language
equally regardless of test-set size, which is appropriate given
the large variance in per-language data volumes.

% ===========================================================
\section{Results}
\label{sec:results}

\subsection{Effect of Encoder and Fine-tuning}
Table~\ref{tab:encoder} compares frozen and fine-tuned settings
for both encoders using CE loss. The two encoders exhibit
contrasting behaviour. For Whisper, fine-tuning yields substantial
gains across all benchmarks (+15.8\% on Vaani-Test, +10.8\% on
FLEURS, +10.6\% on Kathbath), indicating that Whisper's
pre-trained representations, while broadly multilingual, benefit
considerably from task-specific adaptation to Indic LID.
For FastConformer, fine-tuning provides only a marginal improvement
on Vaani-Test (+0.2\%) but noticeably \emph{degrades} cross-corpus
performance (FLEURS: 94.2\% $\to$ 89.9\%; Kathbath: 90.9\% $\to$
87.4\%), suggesting that the frozen FastConformer encoder generalizes
better out-of-domain, and that fine-tuning on Vaani causes it to
overfit to in-domain acoustic conditions. Notably, the frozen
FastConformer outperforms the fine-tuned Whisper on FLEURS and Kathbath, highlighting the strong cross-lingual transferability of FastConformer's pre-trained representations for
Indic languages.

\begin{table}[t]
  \caption{Macro accuracy (\%) for frozen vs.\ fine-tuned encoders
           with CE loss. \textbf{Bold} = best per encoder.}
  \label{tab:encoder}
  \centering
  \begin{tabular}{llccc}
    \toprule
    \textbf{Encoder} & \textbf{Setting}
      & \textbf{Vaani-Test} & \textbf{FLEURS} & \textbf{Kathbath} \\
    \midrule
    \multirow{2}{*}{Whisper}
      & Frozen      & 56.0 & 61.9 & 57.7 \\
      & Fine-tuned  & \textbf{71.8} & \textbf{72.7} & \textbf{68.3} \\
    \midrule
    \multirow{2}{*}{FastConformer}
      & Frozen      & 67.4 & \textbf{94.2} & \textbf{90.9} \\
      & Fine-tuned  & \textbf{67.6} & 89.9 & 87.4 \\
    \bottomrule
  \end{tabular}
\end{table}

\subsection{Effect of Encoder and Training Objective}
Table~\ref{tab:loss} reports macro accuracy for both fine-tuned
encoders across all three training objectives, alongside two
external LID baselines: Facebook MMS(FBMMS)~\cite{pratap2023mms} and
SpeechBrain ECAPA-TDNN(Speechbrain) \cite{valk2021slt},\cite{desplanques2020ecapa}. MMS covers only 30 of
the 42 target languages and ECAPA covers 13, so their Vaani-Test
scores are not directly comparable to our systems.

HSM consistently outperforms CE and CE+SupCon for both encoders
across all benchmarks. For Whisper, HSM achieves the best results
(Vaani-Test: 74.2\%, FLEURS: 73.8\%, Kathbath: 75.8\%), with the
largest gain on Kathbath (+7.5\% over CE), suggesting that the
hierarchical structure aids out-of-domain generalization. For
FastConformer, HSM similarly yields the best results across all
benchmarks. CE+SupCon underperforms CE for FastConformer, suggesting
the contrastive objective interferes with its strong pre-trained
cross-lingual representations.

Among the external baselines, FastConformer + HSM matches SpeechBrain
ECAPA on FLEURS (91.4\% vs.\ 91.3\%) and outperforms it on
Kathbath (90.0\% vs.\ 87.9\%) despite covering over three times
as many languages. MMS achieves 95.1\% on FLEURS but only over
its supported 30 languages; its low Vaani-Test score (33.7\%)
reflects the many Vaani-specific low-resource languages absent
from its inventory.

\begin{table}[t]
  \caption{Macro accuracy(\%) across test benchmarks for each
           training objective.\textbf{Bold} = best per benchmark.}
  \label{tab:loss}
  \centering
\begin{tabular}{llccc}
  \toprule
  \textbf{Model} & \textbf{Objective}
    & \textbf{Vaani-Test} & \textbf{FLEURS} & \textbf{Kathbath} \\
  \midrule
  \multirow{3}{*}{Whisper}
    & CE                 & 71.8          & 72.7             & 68.3 \\
    & CE+SupCon        & 72.4          & 68.1             & 71.0 \\
    & HSM                & \textbf{74.2} & \textbf{73.8}    & \textbf{75.8} \\
  \midrule
  \multirow{3}{*}{FastConformer}
    & CE                 & 67.6          & 89.9             & 87.4 \\
    & CE+SupCon        & 64.5          & 82.7             & 79.8 \\
    & HSM                & \textbf{67.7} & \textbf{91.4}    & \textbf{90.0} \\
  \midrule
  FBMMS          & ---            & 33.7$^\dagger$   & 95.1             & 91.3 \\
  Speechbrain     & ---            & 65.1$^\ddagger$  & 91.3             & 87.9 \\
  \bottomrule
  \multicolumn{5}{l}{\footnotesize $^\dagger$ 30/42 languages supported \quad $^\ddagger$ 13/42 languages supported}
\end{tabular}
\end{table}

\subsection{Cross-corpus Generalization}

A consistent pattern across both tables is that FastConformer
generalizes substantially better to out-of-domain benchmarks
than Whisper, regardless of training objective. The frozen
FastConformer achieves 94.2\% on FLEURS and 90.9\% on Kathbath,
compared to 61.9\% and 57.7\% for the frozen Whisper. Even after fine-tuning, FastConformer with HSM (FLEURS: 91.4\%, Kathbath: 90.0\%) far outperforms the best Whisper result (FLEURS: 73.8\%, Kathbath: 75.8\%). This suggests that FastConformer's pre-training produces more domain-robust acoustic representations for Indic languages, while Whisper's representations are more sensitive to domain shift despite its larger pre-training corpus. The CE+SupCon objective degrades cross-corpus performance for FastConformer (FLEURS: 82.7\% vs.\ 89.9\% for CE), indicating
that the contrastive loss may over-specialize the embedding space
to Vaani's acoustic conditions.

% ===========================================================
\section{Analysis}
\label{sec:analysis}

\subsection{Per-family Performance}

Table~\ref{tab:family} breaks down macro accuracy by language
family and sub-family on Vaani-Test set for both encoders with HSM loss. The results reveal a striking contrast between language groups. Sino-Tibetan languages achieve the highest accuracy
(97.1\%), which is somewhat surprising given the limited training
data for languages such as Wancho, Garo, and Chakma; this likely
reflects that these languages are acoustically highly distinct
from the Indo-Aryan and Dravidian majority, making them easier
to separate despite fewer training samples. English similarly
achieves 92.3\%, consistent with it being the only European
language and therefore trivially distinct from all others.
Dravidian languages also perform strongly at 85.9\%, reflecting
the relative phonological distinctiveness of Tamil, Telugu,
Kannada, and Malayalam from one another and from Indo-Aryan
varieties.
 
Indo-Aryan languages as a group are the most challenging (67.6\%),
and the sub-family breakdown reveals where the difficulty lies.
Central Indo-Aryan is by far the hardest sub-family at 58.7\%,
comprising closely related varieties such as Hindi, Bhojpuri,
Chhattisgarhi, Sadri, Surgujia, Bajjika, and Halbi that share
substantial phonological and lexical overlap. Eastern (84.8\%),
Western (76.8\%), and Northern (79.9\%) Indo-Aryan sub-families
perform considerably better, likely because languages within
these groups are somewhat more acoustically distinct from each
other, or because they have less within-group overlap in the
training data. The low Central Indo-Aryan accuracy is the primary
driver of the overall Indo-Aryan deficit and represents the key
remaining challenge for large-scale Indic LID.
\begin{table}[!t]
  \caption{Per-family macro accuracy (\%) on Vaani-Test
           for both encoders with HSM loss.}
  \label{tab:family}
  \centering
  \begin{tabular}{lcc}
    \toprule
    \textbf{Language Family / Sub-family}
      & \textbf{Whisper} & \textbf{FastConformer} \\
    \midrule
    Indo-Aryan (all)     & 67.6 & 58.9 \\
    \quad Central        & 58.7 & 47.5 \\
    \quad Eastern        & 84.8 & 88.0 \\
    \quad Western        & 76.8 & 75.9 \\
    \quad Northern       & 79.9 & 56.0 \\
    Dravidian            & 85.9 & 84.3 \\
    Sino-Tibetan         & 97.1 & 95.4 \\
    European (English)   & 92.3 & 92.2 \\
    \midrule
    \textbf{Overall}     & \textbf{74.2} & 67.7 \\
    \bottomrule
  \end{tabular}
\end{table}

\subsection{Confusion Among Related Languages} 
\begin{figure}[!t]
  \centering
  \includegraphics[width=\columnwidth]{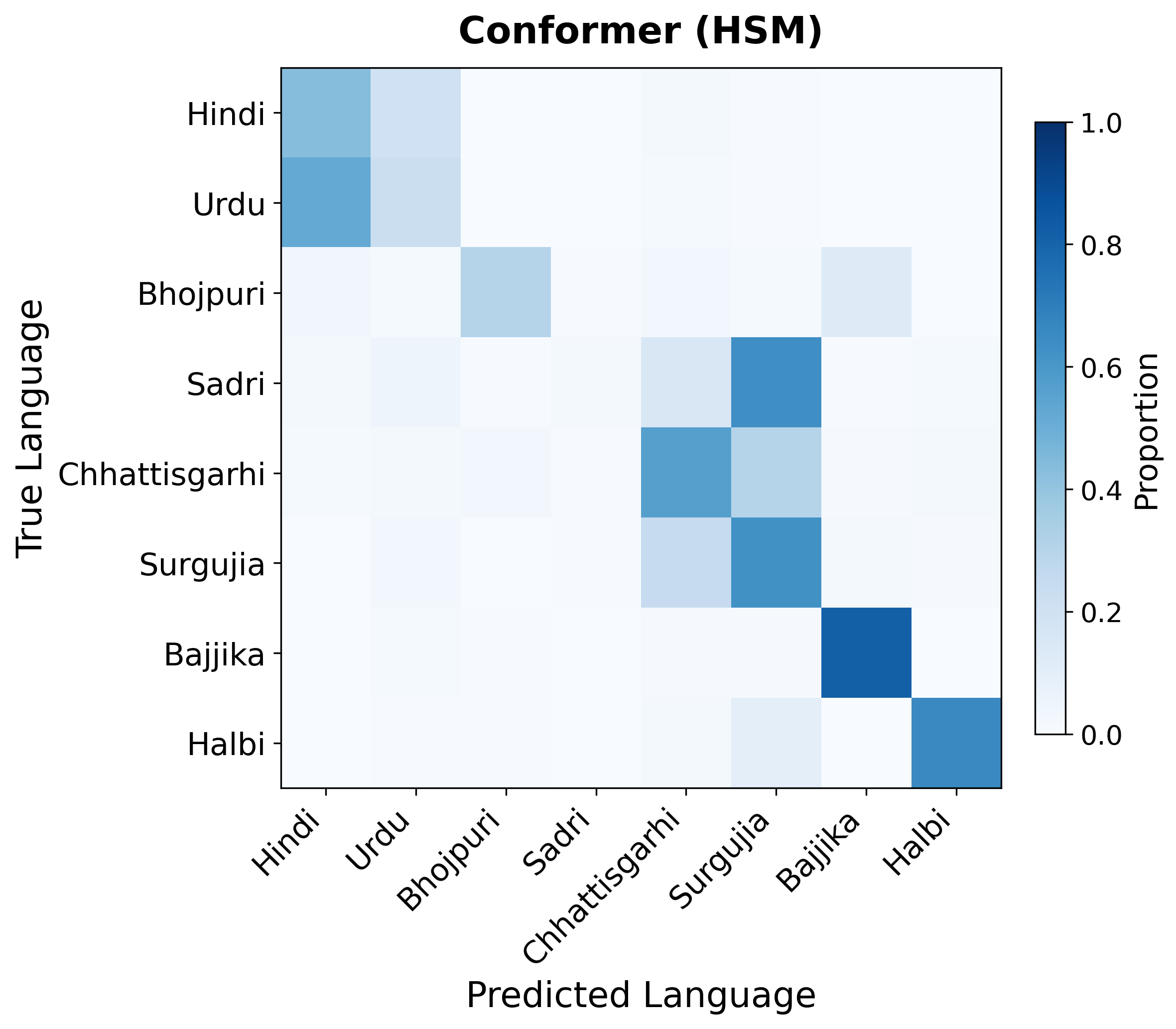}
    \includegraphics[width=\columnwidth]{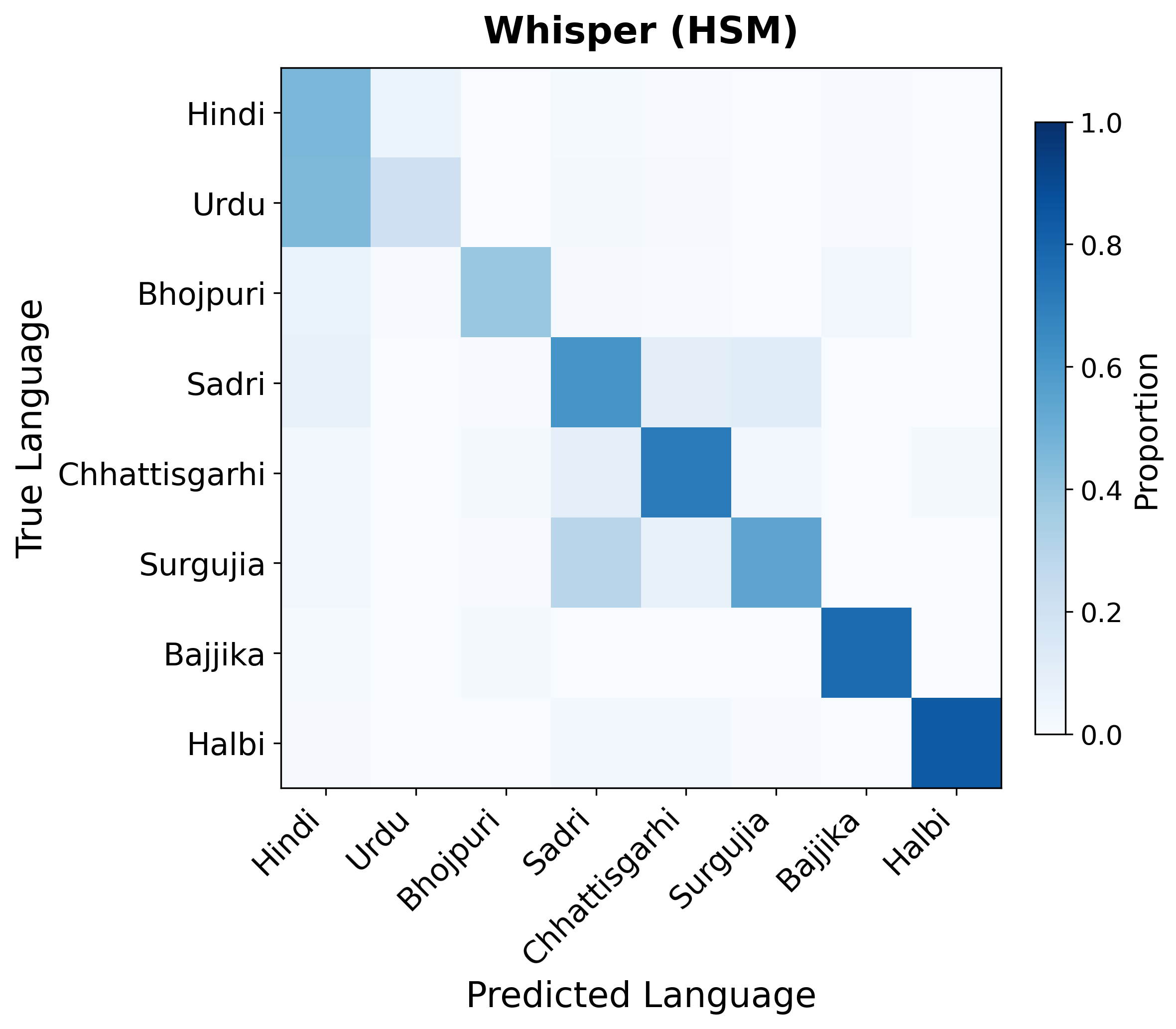}

  \caption{Confusion matrices for Central Indo-Aryan languages: Whisper + HSM  vs.\ FastConformer + HSM. (Darker diagonal = better discrimination.)}
  \label{fig:confmat}
\end{figure}

Fig.~\ref{fig:confmat} shows confusion matrices for Central
Indo-Aryan languages under the HSM objective for both encoders.
Two dominant confusion patterns emerge across both models. First,
Hindi and Urdu are heavily mutually confused, which is linguistically
expected given their near-identical spoken form --- both share the
same phonological inventory and differ primarily in script and
formal vocabulary. This confusion is more severe in FastConformer,
where Urdu is almost entirely misclassified as Hindi, whereas
Whisper maintains a somewhat stronger Urdu diagonal.
 
Second, the Sadri--Chhattisgarhi--Surgujia cluster exhibits
significant mutual confusion in both encoders, reflecting the
high degree of phonological overlap among these closely related
Central Indo-Aryan varieties. In the FastConformer matrix, Sadri
shows weak diagonal activation with mass spread across
Chhattisgarhi and Surgujia. Whisper handles this cluster somewhat
better, producing more distinct diagonal entries for Sadri and
Chhattisgarhi. In contrast, Bajjika and Halbi are well-identified
by both encoders, likely due to sufficient and acoustically
distinctive training samples for these languages. Bhojpuri also
shows moderate accuracy in both models, though FastConformer exhibits
some confusion between Bhojpuri and Bajjika.

% ===========================================================
\section{Conclusion}
\label{sec:conclusion}

We presented a systematic comparison of pre-trained speech encoder
architectures and training objectives for large-scale Indic spoken
language identification across 42 languages. Our cross-corpus
evaluation on Vaani-Test, FLEURS, and Kathbath shows that the
FastConformer encoder substantially outperforms Whisper on
out-of-domain benchmarks, with the frozen FastConformer achieving
over 90\% macro accuracy on FLEURS and Kathbath without any
task-specific fine-tuning, while fine-tuning Whisper yields
stronger in-domain performance. Among training objectives, HSM
consistently outperforms both CE and CE+SupCon for both encoders
across all benchmarks, with the most pronounced gains on Kathbath.
CE+SupCon degrades FastConformer's cross-corpus generalization,
suggesting that the contrastive objective over-specializes
representations to in-domain acoustic conditions. Per-family
analysis reveals that Central Indo-Aryan varieties are the hardest
to discriminate (58.7\%), with Hindi--Urdu and the
Sadri--Chhattisgarhi--Surgujia cluster accounting for the majority
of errors, owing to their high degree of phonological overlap.
Sino-Tibetan and Dravidian languages are comparatively
well-identified despite differing resource levels, highlighting
the importance of inter-language acoustic distinctiveness.
Future work will explore targeted data augmentation for
low-accuracy Central Indo-Aryan varieties and language-family-aware
training strategies.

% ===========================================================
% Acknowledgment — uncomment for camera-ready only
% \section*{Acknowledgment}
% The authors thank [TODO] for compute support.

% ===========================================================
\bibliographystyle{IEEEtran}
\bibliography{references}

\section{Appendix}
\subsection{Languages}
\label{app:languages}
We consider 42 languages organized into a hierarchical taxonomy spanning multiple language families. The languages were chosen on the basis of sufficient durations being available. The Indo-Aryan family is divided into four sub-groups: Central (Hindi, Urdu, Bhojpuri, Sadri, Chhattisgarhi, Surgujia, Bajjika, Halbi), 
Eastern (Angika, Bengali, Assamese, Maithili, Magahi, Khortha, Odia, Sambalpuri, Surjapuri, Nepali, Nagamese), 
Western (Marathi, Konkani, Rajasthani, Marwari, Gujarati, Malvani), and 
Northern (Punjabi, Kumaoni, Garhwali, Haryanvi, Khariboli). 
The Dravidian family includes Telugu, Kannada, Tamil, Tulu, and Malayalam. 
The Sino-Tibetan family comprises Wancho, Sumi, Garo, Kokborok, and Chakma, while the European family is represented by English.

\end{document}